\documentclass[aps,prapplied,reprint,superscriptaddress,longbibliography,amsmath,amssymb, twocolumn]{revtex4}

\usepackage[utf8]{inputenc}
\usepackage{microtype}
\usepackage[T1]{fontenc}

\usepackage{hyperref}
\usepackage{enumitem}
\usepackage{physics}
\usepackage{graphicx, import}

\begin{document}

\title{Local nanoscale probing of electron spins using NV centers in diamond}

\author{Sergei Trofimov}
\affiliation{Berlin Joint EPR Laboratory and Department Spins in Energy Conversion and Quantum Information Science (ASPIN), Helmholtz-Zentrum Berlin für Materialien und Energie, Hahn-Meitner-Platz 1, 14109 Berlin, Germany}

\author{Christos Thessalonikios}
\affiliation{Berlin Joint EPR Laboratory and Department Spins in Energy Conversion and Quantum Information Science (ASPIN), Helmholtz-Zentrum Berlin für Materialien und Energie, Hahn-Meitner-Platz 1, 14109 Berlin, Germany}

\author{Victor Deinhart}
\affiliation{Leibniz-Institut für Höchstfrequenztechnik, Ferdinand-Braun-Institut (FBH), 12489 Berlin, Germany}
\affiliation{Max Born Institute for Nonlinear Optics and Short Pulse Spectroscopy, 12489 Berlin, Germany}

\author{Alexander Spyrantis}
\affiliation{Leibniz-Institut für Höchstfrequenztechnik, Ferdinand-Braun-Institut (FBH), 12489 Berlin, Germany}

\author{Lucas Tsunaki}
\affiliation{Berlin Joint EPR Laboratory and Department Spins in Energy Conversion and Quantum Information Science (ASPIN), Helmholtz-Zentrum Berlin für Materialien und Energie, Hahn-Meitner-Platz 1, 14109 Berlin, Germany}

\author{Kseniia Volkova}
\affiliation{Berlin Joint EPR Laboratory and Department Spins in Energy Conversion and Quantum Information Science (ASPIN), Helmholtz-Zentrum Berlin für Materialien und Energie, Hahn-Meitner-Platz 1, 14109 Berlin, Germany}

\author{Katja Höflich}
\affiliation{Leibniz-Institut für Höchstfrequenztechnik, Ferdinand-Braun-Institut (FBH), 12489 Berlin, Germany}

\author{Boris Naydenov}
\email{boris.naydenov@helmholtz-berlin.de}
\affiliation{Berlin Joint EPR Laboratory and Department Spins in Energy Conversion and Quantum Information Science (ASPIN), Helmholtz-Zentrum Berlin für Materialien und Energie, Hahn-Meitner-Platz 1, 14109 Berlin, Germany}

\begin{abstract}
  Substitutional nitrogen atoms in a diamond crystal (P1 centers) are, on one hand, a resource for creation of nitrogen-vacancy (NV) centers, that have been widely employed as nanoscale quantum sensors. On the other hand, P1's electron spin is a source of paramagnetic noise that degrades the NV's performance by shortening its coherence time. Accurate quantification of nitrogen concentration is therefore essential for optimizing diamond-based quantum devices. However, bulk characterization methods based on optical absorption or electron paramagnetic resonance often overlook local variations in nitrogen content. In this work, we use a helium ion microscope to fabricate nanoscale NV center ensembles at predefined sites in a diamond crystal containing low concentrations of nitrogen. We then utilize these NV-based probes to measure the local nitrogen concentration on the level of 230~ppb (atomic parts per billion) using the double electron-electron resonance (DEER) technique. Moreover, by comparing the DEER spectra with numerical simulations, we managed to determine the concentration of other unknown paramagnetic defects created during the ion implantation, reaching 15~ppb depending on the implantation dose.
\end{abstract}

\maketitle


\section{Introduction}\label{sec:intro}

Nitrogen is one of the most widespread impurities in natural and artificial diamond~\cite{ashfold_nitrogen_2020}. While being essential for creation of nitrogen-vacancy (NV) centers which are extensively used as quantum sensors~\cite{taylor_high-sensitivity_2008,dolde_electric-field_2011,neumann_high-precision_2013}, excessive substitutional nitrogen concentration induces magnetic noise, leading to decoherence of the NV spin states~\cite{hanson_room-temperature_2006}. It is especially difficult to reduce the nitrogen content in crystals grown by the high pressure high temperature method (HPHT), where typical nitrogen concentration in the purest crystals is on the order of 1 ppm (atomic parts per million), which was shown to already significantly influence the coherence time of NV centers~\cite{bauch_decoherence_2020,park_decoherence_2022}. For optimization of the diamond growth process aimed at decreasing the impurity content, one needs a reliable method for measuring the defect concentration. In case of HPHT diamond, the nitrogen content can locally vary by an order of magnitude in the same crystal~\cite{burns_growth-sector_1990}, thus making a spatially resolving detection method invaluable.

Typically, the substitutional nitrogen concentration in diamond is measured with infrared absorption (C center absorption spectrum) and if it is not detectable, the crystal is referred to as IIa type~\cite{zaitsev_optical_2001}. However, the detection limit of this method is on the order of 1 ppm~\cite{hainschwang_origin_2013}, which is still high for quantum sensing applications. Lower nitrogen concentrations can be probed using absorption in the ultraviolet spectral range with the detection limit of about 100~ppb (atomic parts per billion)~\cite{sumiya_high-pressure_1996,de_weerdt_determination_2008,luo_rapid_2022}. Both methods have a spatial resolution on the order of tens of microns, making it difficult to study local variations in the nitrogen concentration~\cite{babich_specifics_2010}.

Nitrogen at a concentration down to the 1~ppb level can be detected with electron paramagnetic resonance (EPR) spectroscopy~\cite{cann_magnetic_2009}, where it is typically referred to as P1 center. However, quantitative measurements are challenging due to its long spin relaxation times~\cite{mitchell_x-band_2013}, the need for a reference sample~\cite{eaton_quantitative_2010}, and the limited spatial resolution.

Double electron-electron resonance (DEER) technique was suggested to overcome the challenges of EPR spectroscopy on nitrogen and enable quantitative bulk measurements of low P1 concentrations without reference samples~\cite{stepanov_determination_2016}.
This method was later developed to a local DEER technique~\cite{rubinas_optical_2021,li_determination_2021}, where P1 centers are probed through NV center ensembles, benefiting from the diffraction limited probe volume due to the laser excitation and photoluminescence (PL) readout of NVs.

The spatial resolution of such a sensor can be further improved using spatially resolved implantation techniques for localized NV center creation by direct writing using a focused ion beam~\cite{hoflich_roadmap_2023}, for example, in a helium ion microscope (HIM)~\cite{huang_diamond_2013,mccloskey_helium_2014}.

Here, we utilize this method to create small ensembles of a few tens of NV centers, which are used for determination of the local defect concentration in a type IIa HPHT diamond crystal using the DEER technique. Since the ion beam in the HIM is focused to a nanometer-sized spot, the effective size of the probed volume is defined by the ion straggle during implantation and the vacancy diffusion during annealing, which allows to improve the spatial resolution beyond the diffraction limit.

We begin in Section~\ref{sec:MandM} with discussing the fabrication of the diamond sample, the experimental setup and the DEER methods utilized in general and in the framework of NVs coupled to P1 centers. In Section~\ref{sec:results}, we present the results of the NV center creation in the studied sample as well as DEER measurements of the paramagnetic defect concentrations and compare them with numerical simulations. Additional information, including auxiliary experimental details and the theoretical framework, is provided in the Supplemental Material (SM).


\section{Materials and Methods}\label{sec:MandM}

\subsection{Diamond Sample}\label{sec:sample}

The sample used in this work is a (001)-oriented 3~mm $\times$ 3~mm $\times$ 0.37~mm single-crystal diamond plate cut out of a type IIa bulk diamond crystal grown under high pressure (5~GPa) and high temperature (1750~K) conditions with a temperature gradient (TG-HPHT)~\cite{trofimov_spatially_2020}. 
The surface of the sample was mechanically polished to a roughness R$_{\text{a}} < 5$~nm.

\begin{figure*}[!ht]
  \centering \includegraphics{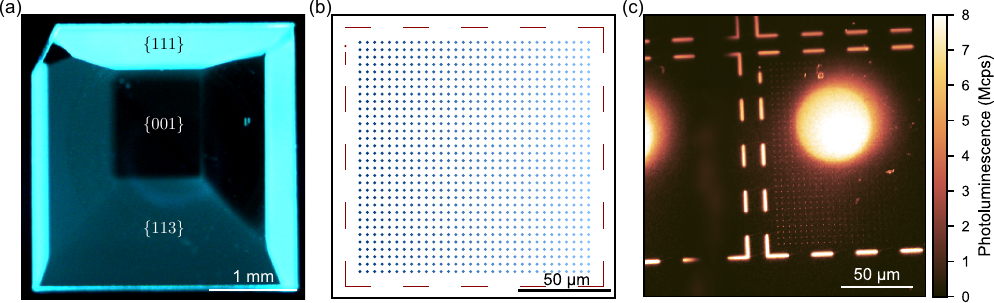}
  \caption{(a) Sample luminescence in UV with an indication of the growth sectors. (b) One out of 9 patterning sites of the implantation pattern used in the helium ion microscope, containing all used ion doses of the other 8 patterning sites. The blue gradient indicates the different doses, where each dose is used for two vertical rows of dots: 5000 (dark blue), 2500, 1000, 500, 250, 100, and from 50 to 5 ions (light blue) in steps of 5 ions. Marker structures are indicated in brown. (c) PL map of the sample obtained in $\{ 111 \}$ GS. The bright circular area is caused by neutralized ions that are not steered and deflected by the electrostatic lens system and become apparent due to the long overall patterning times~\cite{hoflich_roadmap_2023}.\label{fig:sample}}
\end{figure*}

The TG-HPHT growth process conditions are similar to the ones in the Earth's mantle, where diamond crystals grow naturally-faceted, due to a different growth rate for each facet. This allows isolating the so-called growth sectors (GS) of diamond~\cite{dhaenens-johansson_synthesis_2022}. Moreover, the growth sectors differ in defect concentration, which can be visualized by luminescence in ultraviolet (UV), as demonstrated in Figure~\ref{fig:sample}~(a). In the image, one can locate in the order of a decreasing nitrogen concentration (according to Ref.~\cite{burns_growth-sector_1990}): the brightest blue $\{111\}$ GS, the black $\{001\}$ GS, and the pale blue $\{113\}$ GS.

The nitrogen concentration in the $\{ 111 \}$ GS is typically on the order of 500~ppb for similar samples, while in the $\{ 001 \}$ GS it is estimated to be $< 50$~ppb~\cite{trofimov_spatially_2020}. The average nitrogen concentration in this sample was estimated using EPR to be $\approx 20$~ppb (see Section~S1 in SM), which corresponds to $\approx 80$~ppb in $\{ 111 \}$ GS under the assumption, that in other GSs the nitrogen content is at least 10 times lower.

After confirming by confocal microscopy that single NV centers could not be detected in the sample, it was subjected to a $^4\text{He}^+$ implantation in a helium ion microscope (Zeiss Orion NanoFab). The implantation was conducted in all 3 GSs of the crystal according to the pattern shown in Figure~\ref{fig:sample}~(b). Each spot there indicates the position where the ion beam dwelled. The ion energy in the process was set to be 30~keV, while the dose varied from 5 to 5000 ions per spot. 
According to simulations conducted using the Stopping and Range of Ions in Matter (SRIM) software~\cite{ziegler_srim_2010}, each ion creates on average 37 vacancies that have a maximum of their depth distribution at $\approx$~120~nm (see Section~S2 in the SM).

After the implantation, the sample was annealed in argon atmosphere at pressure $P = 7$~mbar using the following steps: 1 hour at 400\textdegree{C}, 1 hour ramping up to 990\textdegree{C}, 2 hours at 990\textdegree{C}, 2 hours cooling down.


\subsection{Experimental methods}\label{sec:methods}

\begin{figure}[!ht]
  \centering \includegraphics{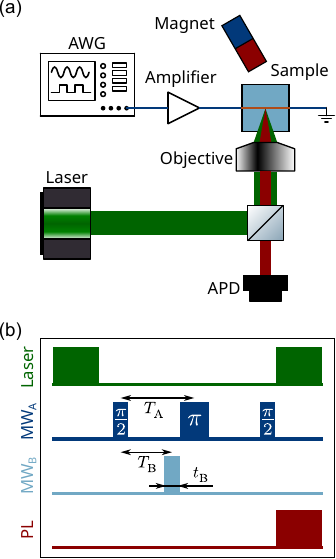}
  \caption{(a) Schematics of the experimental setup. MW pulses for spins A and B were generated by an AWG. A permanent magnet was used to create a static magnetic field. NV centers were excited with a green laser focused by an air objective. The same objective was used to collect the NV center PL that was measured by an APD. (b) Pulse sequence used for DEER experiments. It represents a standard Hahn echo pulse sequence used for NV centers with an additional MW pulse for driving spin transitions in target defects.\label{fig:setup}}
  
\end{figure}

All NV center measurements were performed on a home-built confocal microscope (see Figure~\ref{fig:setup}~(a)) described in detail in~\cite{volkova_glovebox-integrated_2025}. During the experiments, a magnetic field of $B_0 \approx 37$~mT created by a permanent magnet was manually aligned with one of the NV center's crystallographic orientation. NV and P1 centers were excited by microwave (MW) signals with frequencies $f^{}_{\text{NV}} \approx 1.8$ GHz and $f^{}_{\text{P}1} \approx 1$~GHz. Both were generated by an arbitrary waveform generator (AWG, AWG7122C, Tektronix) and after amplification (ZVA-213-S+, Mini-Circuits) applied to the sample via a 20~\textmu{m} thick copper wire. NV centers were optically excited with a green continuous wave (CW) laser (iBEAM-SMART-515-S, Toptica Photonics) focused through an air objective (0.95 NA, MPLAPON, Olympus). The NV photoluminescence was collected with the same objective and guided to an avalanche photodiode (APD, SPCM-AQRH-14, Excelitas). Time-correlated single photon counting was done using a fast counter (MCS6, FASTComTec).

For controlling the experiments and data acquisition, the Qudi software suite was used~\cite{binder_qudi_2017}. Numerical simulations of pulsed experiments were done using the Quantum Color Center Analysis Toolbox (QuaCCAToo) software~\cite{quaccatoo_paper, tsunaki_quantum_2024,tsunaki_ambiguous_2025}. Monte-Carlo numerical simulations of the helium ion implantation were conducted using SRIM~\cite{ziegler_srim_2010}.

EPR measurements were performed on a CW-EPR spectrometer (MS 5000, Magnettech).

The visualization of GSs of the sample using luminescence in UV was done with a DiamondView device (De Beers).


\subsection{Double electron-electron resonance technique}\label{sec:deer}

In DEER method (also known as pulsed electron double resonance, PELDOR), target spins B (in our case P1 centers and other defects) are probed through sensor spins A (NV centers)~\cite{milov_pulsed_1998}. The measurement protocol demonstrated in Figure~\ref{fig:setup}~(b) consists of a standard Hahn echo pulse sequence~\cite{hahn_spin_1950} (the last $\pi/2$-pulse is needed for the optical readout of NV centers) with a constant delay time $T_\text{A}$ applied to spins A and an additional DEER MW pulse inducing spin transitions in the target defects. When the frequency of the DEER pulse $f^{}_{\text{B}}$ is equal to the target spin transition frequency, the final state of spins A accumulates an additional phase shift $\phi^{}_\text{DEER}$, which in case of NV centers results in a change of the luminescence at the readout (for more details see Appendix~\ref{app:deer_nv}). Apart from the frequency $f^{}_\text{B}$, the DEER signal $I^{}_\text{DEER}$ depends on the duration of the DEER MW pulse $t^{}_\text{B}$ and on the delay time $T^{}_\text{B}$ between the first MW $\pi/2$-pulse and the DEER pulse according to the equation~\cite{stepanov_determination_2016,li_determination_2021}
\begin{gather}
  I^{}_\text{DEER}(f^{}_\text{B}, t^{}_\text{B}, T^{}_\text{B}) = \notag \\
  \exp[-\frac{4 \pi \mu^{}_0 \mu^{2}_\text{B} g^{}_\text{A} g^{}_\text{B} |\sigma^{}_\text{B}| n^{}_\text{B}}{9 \sqrt{3} \hbar} T^{}_\text{B} P^{}_\text{B}(f^{}_\text{B}, t^{}_\text{B})],
  \label{eq:deer}
\end{gather}
where $\mu^{}_0$ is the vacuum permeability, $\mu^{}_\text{B}$ is the Bohr magneton, $g^{}_\text{A}$ and $g^{}_\text{B}$ are the sensor and target spin g-factors respectively, $n^{}_\text{B}$ is the target defect concentration, $\hbar$ is the reduced Planck constant, and $ |\sigma_\text{B}| = \frac{1}{2} \left| \expval{\mathcal{S}_z}{1}_\text{B} - \expval{\mathcal{S}_z}{2}_\text{B} \right|$ for the $\ket{1}_\text{B} \leftrightarrow \ket{2}_\text{B} $ transition of the target electron spin. $P^{}_\text{B}(f^{}_\text{B}, t^{}_\text{B})$ is the effective population transfer which represents a convolution of the target spin EPR spectrum $L$ with the transition probability $P^{}_\text{R}$, so that $P^{}_\text{B}(f^{}_\text{B}, t^{}_\text{B}) = (L * P^{}_\text{R}) (f^{}_\text{B}, t^{}_\text{B}) = \int_{-\infty}^{\infty} L(\xi) \, P^{}_\text{R} (f^{}_\text{B} - \xi, t^{}_\text{B}) \mathrm{d} \xi$. In this work, we considered the target EPR spectrum to be a normalized sum of Lorentzian functions 
\begin{gather}
  L(\xi) = \sum_{i} \frac{A^{}_i}{\pi} \frac{\Gamma^{}_i}{\Gamma^2_i + {\left( \xi - f^\text{R}_i \right)}^2},
  \label{eq:lorentz}
\end{gather}
where $f^\text{R}_i$, $\Gamma_i$ and $A_i$ are the resonance frequency, the half-width at half-maximum, and the amplitude of the $i$-th resonance in the EPR spectrum of target defects correspondingly. The sum of the amplitudes is normalized to unity: $\sum_{i} A^{}_i = 1$. The transition probability $P^{}_\text{R} (f^{}_\text{B} - \xi, t^{}_\text{B})$ is obtained using the Rabi formula
\begin{gather}
  P^{}_\text{R} (f^{}_\text{B} - \xi, t^{}_\text{B}) = \notag \\
  \frac{\Omega^2}{\Omega^2 + {\left( f^{}_\text{B} - \xi \right)}^2} \sin^2 \left[ 2 \pi \sqrt{\Omega^2 + {\left( f^{}_\text{B} - \xi \right)}^2} \frac{t^{}_\text{B}}{2} \right],
  \label{eq:rabi}
\end{gather}
where $\Omega$ is the Rabi frequency of spins B.

By measuring the DEER signal as a function of the excitation frequency $I^{}_\text{DEER}(f^{}_\text{B})$ one obtains a DEER spectrum. DEER Rabi oscillations are obtained by measuring the dependency on the length of the DEER pulse $I^{}_\text{DEER}(t^{}_\text{B})$ for each of the target spin resonances in the DEER spectrum. Finally, a DEER decay $I^{}_\text{DEER}(T^{}_\text{B})$ is probed when the delay time $T^{}_\text{B}$ is varied. Fitting these experimental data sets with Equation~\ref{eq:deer} allows to calculate the local target defect concentration $n^{}_\text{B}$.


\subsection{DEER signal from P1 and NV centers}\label{sec:deer_on_p1}

The determination of the target spin concentration relies on measuring the NV spin-dependent PL in a Hahn echo pulse sequence, while the spin state of the target defect is manipulated via the DEER pulse (see Appendix~\ref{app:deer_nv} for more details).

To enhance the spin contrast of the detected signal, the external magnetic field is typically aligned with NV centers from one of the $\langle 111 \rangle$ crystallographic orientation families in diamond, which are then used as sensor spins. The other NV orientation families constitute an angle of $109.5^{\circ}$ with the magnetic field and are referred to as off-axis NVs. In this field configuration, the DEER spectrum of P1 ensembles consists of five resonance peaks (electron spin $S = 1/2$, nuclear spin $I = 1$ for $^{14}$N, and two families of crystallographic orientation) instead of 12 peaks that are present otherwise (electron spin $S = 1/2$, nuclear spin $I = 1$ for $^{14}$N, and four different crystallographic orientations)~\cite{degen_entanglement_2021}.
In this work, we excluded the central peak in the P1 DEER spectrum from the nitrogen concentration analysis, as other electron spin species with the same g-factor $g \approx 2$ can also contribute to the signal at this frequency (see below). Therefore, we used it to determine the concentration of additional defects X. In the analysis we used $g^{}_\text{NV} = g^{}_\text{P1} = g^{}_\text{X} = 2$ and $ |\sigma^{}_\text{P1}| = |\sigma^{}_\text{X}| = 1/2$.

As shown in previous studies, spin transitions of the NV centers that are not parallel to the magnetic field (off-axis NVs) also appear in the DEER spectrum~\cite{li_determination_2021}. Moreover, due to the spin state mixing, transitions between all three spin sublevels $\ket{1},\ \ket{2},\ \ket{3}$ (corresponding to $\ket{m_s = 0},\ \ket{m_s = -1},\ \ket{m_s = +1}$ at zero perpendicular magnetic field) can be excited. In this work we used the $ \ket{2} \leftrightarrow \ket{3} $ NV transition (see details in Appendix~\ref{app:simulations}), where $|\sigma^{}_\text{NV}| = 0.87$ was calculated using numerical simulations. As the NV centers are polarized by the laser, one has to take into account the population of the spin levels, since they cannot be assumed equal (as was implied for P1 centers). Following the analysis carried out in~\cite{li_determination_2021}, we calculated the populations of 40\%, 30\%, and 30\% for the spin levels $ \ket{1} $, $ \ket{2} $, and $ \ket{3} $ respectively. Therefore, the observed transition in DEER originates from 60\% of the off-axis NV centers, or 45\% of all NVs. The details of these calculations are provided in Appendix~\ref{app:simulations}.

To determine the target spin concentration $n^{}_\text{P1}$ one can use all dependencies of $I^{}_\text{DEER}$ on $f^{}_\text{B}$, $t^{}_\text{B}$ or $T^{}_\text{B}$ as described in Section~\ref{sec:deer}. However, due to the high number of free fitting parameters ($n^{}_\text{B}$, $\Gamma^{}_{i}$, $f^\text{R}_i$, $\Omega$) and considering the complex structure of the fitting function, we use the following protocol for the defect concentration determination.

First, a DEER spectrum $I^\text{exp}_\text{DEER}(f^{}_\text{B})$ is measured while sweeping the excitation frequency at $t^{}_\text{B}$ taken equal to the NV center $\pi$-pulse length and at $T^{}_\text{B} = T^{}_\text{A}$. Fitting the spectrum with Lorentzian functions allows to determine the target spin resonance peak positions $f^\text{R}_i$. 

Next, DEER Rabi oscillations $I^\text{exp}_\text{DEER}(t^{}_\text{B})$ are obtained at fixed $f^{}_\text{B} = f^\text{R}_i$ and $T^{}_\text{B}$ to determine the Rabi frequency $\Omega$ and the corresponding correct length of the DEER $\pi$-pulse $ t^{}_{\pi} = 1/(2 \Omega)$. The Rabi frequency $\Omega$ must be measured for each of the resonance peaks separately, since the MW power that reaches the NVs might differ depending on the excitation frequency.

Finally, another DEER spectrum $I^\text{exp}_\text{DEER}(f^{}_\text{B})$ around the resonance frequency $f^\text{R}_i$ is obtained at fixed $t^{}_\text{B} = t^{}_{\pi}$ and $T^{}_\text{B}$. The resulting spectrum is fitted with Equation~\ref{eq:deer} with the parameters and $\Omega$ fixed, while treating $f^\text{R}_i$, $\Gamma^{}_{i}$ and $n^{}_\text{B}$ as free parameters. As stated previously, the normalized amplitude $A^{}_{i}$ is taken from the simulated spectrum for P1 centers or from calculations of the spin level populations for NV centers. For the P1 defects, the final concentration $n^{}_\text{P1}$ is calculated as an average of the values obtained using four resonance peaks (excluding the central transition).


\section{Results and discussion}\label{sec:results}

\subsection{NV center creation and characterization}\label{sec:nv_creation}

Figure~\ref{fig:sample}~(c) shows a PL image obtained in the $\{ 111 \}$ GS showing bright spots with NV centers corresponding to the implantation pattern. However, the PL map also contains a bright circular feature present in all implantation sites with a diameter of about 60~\textmu{m}. This feature is attributed to an unintended implantation of unfocused neutral atoms, which requires a specialized column design to avoid~\cite{hoflich_roadmap_2023}.

\begin{figure*}[!ht]
  \centering \includegraphics{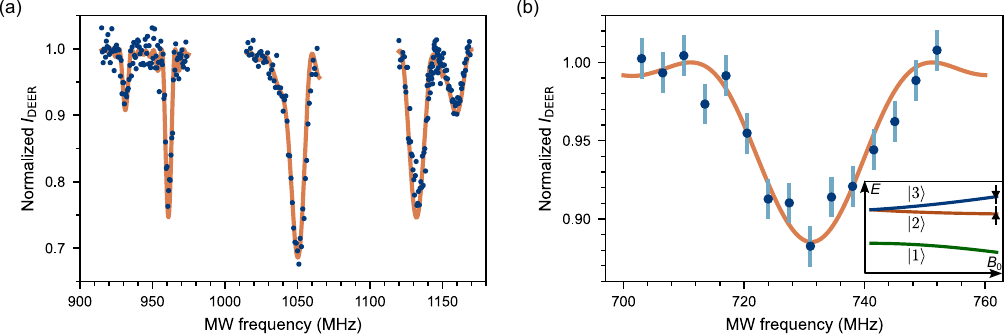}
  \caption{Experimentally measured (blue points) and numerically simulated using QuaCCAToo software~\cite{quaccatoo_paper, tsunaki_quantum_2024} DEER spectra (brown line). The magnetic field used in simulation was 37.2~mT, as calculated from the NV center ODMR spectrum applied at 0.1\textdegree{} to the [111] direction in diamond lattice (for agreement with experimental NV transition in DEER spectra). (a) Nitrogen DEER spectrum at implantation dose 5000 ions/spot. The center transition has a shoulder on its left side, which is attributed to the implantation-related defect X (see main text). The P1 and X concentration used for the simulated spectrum were taken to be 200~ppb and 13~ppb, respectively. (b) DEER spectrum showing the $ \ket{2} \leftrightarrow \ket{3} $ transition of off-axis NVs measured at implantation dose 2500 ions/spot. The 0.1\textdegree{} angle misalignment between the magnetic field and one of the $\langle 111 \rangle$ diamond lattice directions was used in simulations to achieve an agreement with the experimental data. Inset shows the spin states of the off-axis NV ground level as a function of the static magnetic field with an indication of the observed transition (see also Figure~\ref{fig:levels}~(a)). Error bars represent the photon shot noise.\label{fig:simulation}}
\end{figure*}

In the $\{ 111 \}$ GS several implantation spots with different doses ranging from 35 to 5000 ions per spot were chosen for determination of the NV coherence time $T^{}_2$ using Hahn echo decay pulse sequence~\cite{hahn_spin_1950}. An example of such a measurement can be found in Section~S3 in the SM.
The resulting average value is $T_2^{ 111 } = 284 \pm 14$~\textmu{s}. 

The measured $T^{}_2$ of above 100~\textmu{s} is close to the limit set by the natural $^{13}$C concentration ($1.1\%$), confirming that the nitrogen content is indeed low, $n^{}_\text{P1} < 1$~ppm~\cite{bauch_decoherence_2020}. Using Equation~\ref{eq:deer} and taking $0.35 \,  T_2 \approx 100$~\textmu{s} as the optimal value for $T_\text{B}$~\cite{rubinas_optimization_2022}, one can estimate the minimum nitrogen concentration that could be detected with DEER method for this sample. Taking $I^{}_\text{DEER} = 0.95$ as the minimum visible DEER signal (5\% contrast) limited by the noise for the central peak in the P1 DEER spectrum, and considering excitation at high Rabi frequency ($\Omega >> \Gamma$, so that $P^{}_\text{B}(f^{}_\text{B}, t^{}_\text{B}) = 1$) we obtain the P1 concentration detection limit to be $ [\text{P}1]_\text{lim} \approx 5$~ppb.

Using PL of the implanted spots as an indication of the number of created NV centers and assuming it is proportional to the nitrogen content, we estimate the ratio of the P1 concentration between the GSs from the analysis of saturation curves, where the fluorescence intensity is measured as function of the excitation laser power $F_\text{NV}(P_\text{las})$. For that, the PL saturation curves were measured on spots with different implantation doses (3 spots for each dose) in all GSs. To estimate the number of NV centers in each implanted spot, additional saturation curves on single NV centers and on background were obtained in each GS.
After removing the background, the saturation curves for implanted spots and for single NVs are fitted with the following function~\cite{siyushev_monolithic_2010,delgado_impact_2025}:
\begin{equation}
  F_\text{NV}(P_\text{las}) = \frac{F_\text{sat} P_\text{las}}{P_\text{sat} + P_\text{las}}
  \label{eq:saturation}
\end{equation}
Here $F_\text{sat}$ parameter is the PL intensity at saturation, $P_\text{sat}$ is the saturation laser power.
The number of NV centers at each implantation dose for all three GSs is then calculated by dividing the intensity at saturation of the NV ensemble by the saturation intensity of a single NV $N^\text{GS, dose}_\text{NV} = F^\text{GS, dose}_\text{sat} / F^\text{GS, single}_\text{sat}$. 
The P1 center concentration ratio was estimated from the average ratio of the NV centers per dose in each GS: $ n_\text{P1}^{ 111 } : n_\text{P1}^{ 113 } : n_\text{P1}^{ 001 }= \langle N_\text{NV}^\text{111, dose} : N_\text{NV}^\text{113, dose} : N_\text{NV}^\text{001, dose} \rangle_\text{dose} \approx 10.1 : 1.3 : 1.0$. The dependencies of the number of NV centers on the implantation dose are shown in Figure~S4b in the SM.


\subsection{P1 concentration}\label{p1_concentration}

The P1 concentration in $\{ 111 \}$ GS is determined using four of the nitrogen peaks (excluding the central one) in DEER spectra measured on several implantation spots with different $^4$He$^+$ implantation doses ranging from 35 to 5000 ions/spot. The obtained spectrum shows an excellent agreement with the numerical simulation as presented in Figure~\ref{fig:simulation}~(a), indicating that linewidths of resonances are defined by Rabi frequencies of P1 (MW power broadening). Considering the measured $T_2$ and accounting for the $^{13}$C modulation (see, for example,~\cite{van_oort_optically_1990}) visible in the Hahn echo decay measurements on a short scale (see Figure~S3b in the SM), we chose the $T_\text{A}$ parameter to be 20~\textmu{s}. 

\begin{figure*}[!ht]
  \centering \includegraphics{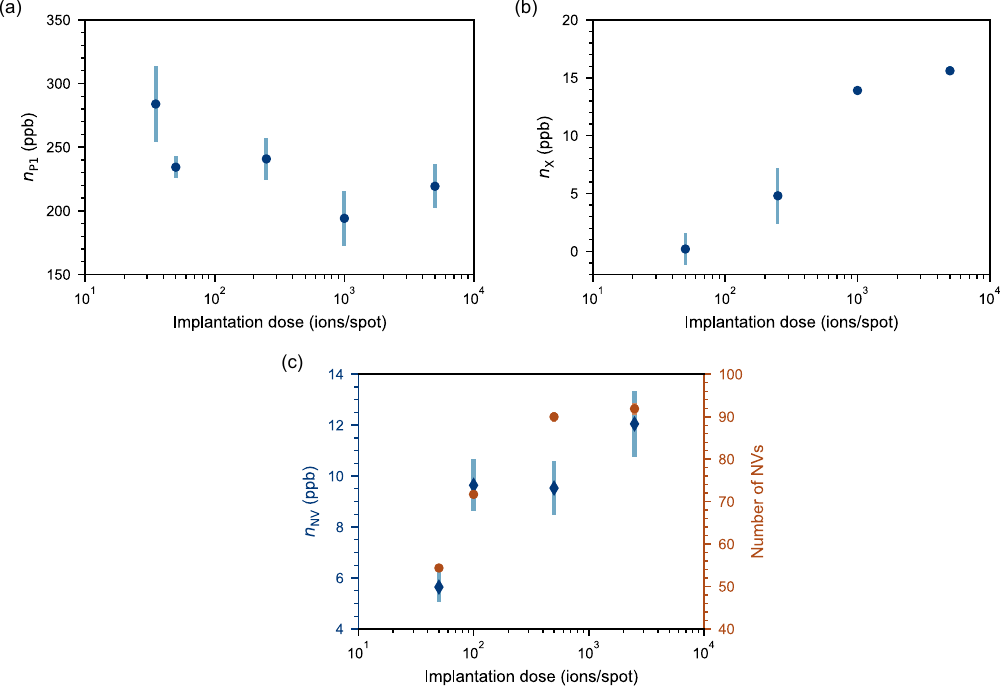}
  \caption{(a) Nitrogen concentration obtained from DEER measurements in $\{ 111 \}$ GS at different implantation doses. Difference in the value of $n^{}_\text{P1}$ for different implantation doses could indicate that P1 centers are being converted to NV centers and other defects and/or by the local nitrogen concentration variations. Error bars represent the standard deviations calculated for $n^{}_\text{P1}$ values obtained on four nitrogen transitions. (b) Concentration of the additional defects obtained from DEER measurements in $\{ 111 \}$ GS at different implantation doses. Increase of $n^{}_\text{X}$ at higher doses allows to attribute the defect origin to ion implantation. (c) NV center concentration (blue diamonds) obtained from DEER measurements and the number of NV centers (brown dots) derived from saturation curves in $\{ 111 \}$ GS at different implantation doses. Error bars represent the standard errors obtained from fitting.\label{fig:nitrogen}}
\end{figure*}

The calculated nitrogen concentration obtained from the DEER spectra is shown in Figure~\ref{fig:nitrogen}~(a) for each measured implantation dose. The dependency of $n^{}_\text{P1}$ on the implantation doses could indicate that P1 centers are more efficiently converted to other defects (NV and X centers) and/or by local variations in the nitrogen content. The averaged P1 concentration $ n_\text{P1}^{ 111 } = 234 \pm 13$~ppb agrees well with the estimated bound given in Section~\ref{sec:sample}. 

As demonstrated in Figure~\ref{fig:simulation}~(a), the experimental data indicate the presence of an additional resonance peak visible as a shoulder of the central nitrogen transition. Assuming that this signal is caused by an additional defect X~\cite{li_determination_2021}, we can calculate its concentration $ n^{}_\text{X} $ by fitting the central transition in each measured DEER spectrum with two peaks using the nitrogen concentration $n^{}_\text{P1}$ determined from the other four transitions. Figure~\ref{fig:nitrogen}~(b) shows the resulting $ n^{}_\text{X}$ as a function of the $^4$He$^+$ implantation dose. While at low doses (50, 250 ions/spot) the additional peak is not visible (fitting therefore results in the amplitude that is within the scattering of the measurement data), the signal from X becomes detectable starting from 1000 ions/spot and increases at 5000 ions/spot. Therefore, we attribute this signal to more complex types of defects created due to the implantation damage. One possibility could be the WAR10 defect representing a pair of an interstitial nitrogen and an interstitial carbon with an electron spin 1/2~\cite{felton_electron_2009}. In Figure~\ref{fig:central_line} we demonstrate that this additional peak can be simulated adding a spin-$1/2$ system. 

Based on the ratios of the P1 concentration given in Section~\ref{sec:nv_creation} in different GSs and on the measured $n_\text{P1}^{ 111 }$, we conclude, that $n_\text{P1}^{ 113 }$ should be on the order of 20~ppb and, therefore, higher than the potential detection limit of 5~ppb. However, we could not detect even the central P1 resonance line in $\{ 113 \}$ GSs. The reason for that could be an overestimated number of NV centers (hence, the nitrogen concentration) derived from saturation curves for $\{ 113 \}$ GS, which did not take into account the presence of other luminescence defects (for example, NV$^0$) in the implanted spots. 
DEER measurements in this GS are further complicated, as the lower number of the created NV centers results in the increased measurement time needed for statistical averaging. 

DEER measurements in $\{ 001 \}$ GS were not conducted, since the estimated number of created NV centers in this GS is approximately the same as in $\{ 113 \}$.

\subsection{NV concentration}

The NV concentration is determined using the $\ket{2} \leftrightarrow \ket{3}$ transition of off-axis NVs in DEER spectra (see Figure~\ref{fig:simulation}~(b)), which were obtained on several spots with different implantation doses ranging from 50 to 2500 ions/spot in $\{ 111 \}$ GS. The $T_\text{A}$ parameter was chosen to be 80~\textmu{s} in order to successfully measure the weak (due to low local NV concentration) NV center DEER signal.

The dependency of the $n_\text{NV}^{ 111 }$ concentration from the $^4 \text{He} ^+$ dose is shown in Figure~\ref{fig:nitrogen}~(c). For comparison, the number of NV centers measured using saturation curves as described in Section~\ref{sec:nv_creation} is also shown. Both dependencies agree qualitatively, showing an increase of the number NV centers at higher implantation doses, as expected.

Dividing the number of NVs by the detected concentration, one can estimate the average volume occupied by NV centers $V^{}_\text{NV} = (4.8 \pm 0.3) \cross 10^{-2}$~\textmu{m}$^3$. Assuming the volume to be spherical for simplicity, we calculate its radius to be $r^{}_\text{NV} = 226 \pm 13$~nm. Taking into account the vacancy distribution obtained from SRIM (the radius of the spherically approximated volume with vacancies $r^{}_\text{vac} = 37.5$~nm) and the diffusion of vacancies during annealing, we derive the diffusion length to be $ \sqrt{ \langle d^2 \rangle} = \sqrt{(r^{}_\text{NV})^2 - (r^{}_\text{vac})^2} = 223 \pm 13$~nm. This allows to calculate the diffusion coefficient $ D_\text{vac} = \langle d^2 \rangle / 6t = 1.2 \pm 0.1$~nm$^2$/s (where $t = 7200$~s is the annealing time), which agrees well with estimates from literature $D_\text{vac}^{1} = 1.8$~nm$^2$/s at $T = 850^{\circ}$C~\cite{alsid_photoluminescence_2019} and $D_\text{vac}^{2} = 1.1$~nm$^2$/s at $T = 750^{\circ}$C~\cite{martin_generation_1999}. 


\section{Conclusion}

In summary, we showed that ion implantation using a focused ion beam of a helium ion microscope enables the creation of NV center ensembles in type IIa HPHT diamond with long coherence times approaching the limit imposed by the natural abundance of $^{13}$C.

We also demonstrated that employing double electron-electron resonance technique allows to locally measure P1 concentration in type IIa diamond crystals with sub-micron spatial resolution. While measurements in $\{ 111 \}$ growth sector of an HPHT sample indicate $234 \pm 13$~ppb of substitutional nitrogen, the P1 DEER signal in other GSs could not be detected, which suggests that the nitrogen concentration there is on the order of the calculated detection limit of 5~ppb.

We showed that the intensity of the additional peak in the DEER spectrum observed previously~\cite{li_determination_2021} increases with the helium ion implantation dose, implying that it is indeed a vacancy-related defect created due to the ion implantation.

The local NV center concentration measured using the DEER technique was demonstrated to also rise with the implantation dose. Combined with the number of NV centers that could be obtained by measuring PL saturation curves or fluorescence autocorrelation function $g^{(2)}(\tau)$, it can be used for studying vacancy diffusion in diamond. The diffusion coefficient obtained here is in good agreement with values reported in the literature. 

The results obtained in this work represent an important milestone in the precise determination of paramagnetic impurities in diamond, with concentrations reaching the atomic parts-per-billion level and at the same providing a spatial resolution below 1~\textmu{m}. Our results contribute to the further development of deterministic and scalable fabrication of diamond-based quantum technology platforms.

\begin{acknowledgments}
The authors thank Dr.\ Sergey A.\ Tarelkin for providing the diamond sample.
This work was supported by German Research Foundation (DFG, grant 410866378) and by German Federal Ministry of Education and Research (BMBF, grant DIAQUAM 13N116956).
\end{acknowledgments}

\appendix
\section{DEER using NV centers}\label{app:deer_nv}
In this work, we used NV centers as quantum sensors for determining the concentration of the surrounding electron spins (spin bath). This method relies on measurements of the NV spin-dependent PL in a Hahn echo pulse sequence, while the spin state of the target defect is manipulated via the second MW pulse (see Figure~\ref{fig:setup}~(b)). Due to limitations of the pulse generation at our experimental setup, the DEER pulse is positioned right before the NV $\pi$-pulse instead of being the middle of the sequence. In our case, this does not affect the results, as the typical length of the DEER $\pi$-pulse in the experiments ($\approx 200$~ns) is negligible compared to the used $T_\text{A}$ (20~\textmu{s} for P1 centers).

\begin{figure}[!ht]
	\centering \includegraphics{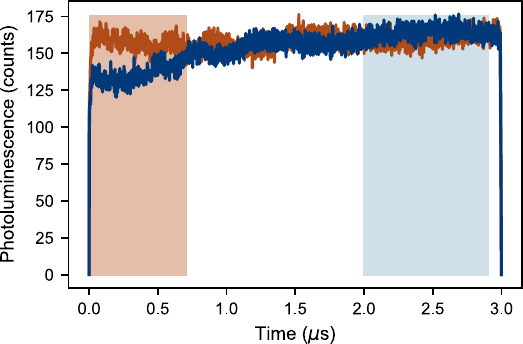}
	\caption{\label{fig:sequence} Illustration of the detection scheme showing the single NV PL during a 3~\textmu{s} laser pulse in the end of the pulse sequence (see Figure~\ref{fig:setup}~(b)). The brown and the blue data correspond to $PL^{\pi/2}_{}$ and $PL^{-\pi/2}_{}$, correspondingly, where the superscript $\pi/2$ ($-\pi/2$) shows the phase of the last MW$_\text{A}$ pulse. The light brown (light blue) region indicate the signal (reference) window with the length of 700~ns (900~ns).} 
\end{figure}

The signal used for the DEER analysis was obtained in an alternating regime, in which two pulse sequences that differ in the phase ($0$ and $\pi$) of the last MW pulse on the NV center spin transition are applied successively. The length of this last MW pulse is the same for both sequences and correspond to the length of the $\pi/2$-pulse. Thus, these pulses correspond to $\pi/2$ and $-\pi/2$ rotation angles in the rotating frame of reference. Below we use superscripts $\pi/2$ and $-\pi/2$ to distinguish between the data obtained using these alternating pulse sequences. For both pulse sequences, the derived signal is defined as a ratio of the NV PL in the signal time window and in the reference time window during the readout laser pulse (see Figure~\ref{fig:sequence}):

\begin{gather}
  \label{eq:pl}
	I^{\pi/2, -\pi/2}_\text{NV} = \frac{PL^{\pi/2, -\pi/2}_{sig}}{PL^{\pi/2, -\pi/2}_{ref}}.
\end{gather}

The resulting NV signal that contains only the coherent evolution of the NV center spin is obtained by subtracting the alternating signals: $I^{}_\text{NV} = \frac{1}{2} \left[ I^{\pi/2}_\text{NV} - I^{-\pi/2}_\text{NV} \right]$. Finally, the experimentally determined DEER signal is defined as the normalized NV signal $I^\text{exp}_\text{DEER} = I^{}_\text{NV}/I^\text{off}_\text{NV}$, where $I^\text{off}_\text{NV}$ is the off-resonant signal obtained at a frequency window $\Delta f^{}_\text{B}$ away from the target spin transitions.

\section{Numerical simulations}\label{app:simulations}
DEER spectra of P1 centers, NV centers, and of the defect X were calculated using QuaCCAToo software~\cite{quaccatoo_paper} which simulates the time evolution of the studied quantum system in the laboratory frame of reference without applying rotating wave approximation~\cite{tsunaki_quantum_2024}. The numerical simulation framework is discussed in detail in Ref.~\cite{tsunaki_ambiguous_2025}.
 To simulate the P1 DEER spectrum we used the following Hamiltonian:
\begin{gather}
	\mathcal{H}_\text{P1} =  \gamma_e \Vec{B}_0 \cdot \Vec{\mathcal{S}} + A_{\perp} \left( \mathcal{S}_x \mathcal{I}_x + \mathcal{S}_y \mathcal{I}_y \right) + A_{\parallel} \mathcal{S}_z \mathcal{I}_z + \notag \\ 
  P_{\parallel}{(\mathcal{I}_z)}^2	- \gamma_{N} \Vec{B}_0 \cdot \Vec{\mathcal{I}},
  \label{eq:p1_h}
\end{gather}
where $\gamma_e = 28.025$~MHz/mT and $\gamma_{N} = 3.077 \times 10^{-3}$~MHz/mT are the electron and nitrogen nucleus gyromagnetic ratios, $\mathcal{S}$ and $ \mathcal{I}$ are the electron and nitrogen nuclear spin operators ($S = 1/2$ and $I = 1$ for $^{14}$N), $\Vec{B}_0$ is the static magnetic field vector, $A_{\perp} = 81.32$~MHz and $A_{\parallel} = 114.03$~MHz are components of the hyperfine interaction between the electron and nuclear magnetic moments, $P_{\parallel} = -3.97$~MHz is the nuclear quadrupole interaction parameter~\cite{cox_13_1994}.

The DEER spectrum of the off-axis NV centers was modeled using the Hamiltonian
\begin{gather}
  \label{eq:nv_h}
	\mathcal{H}_\text{NV} = D {\mathcal{S}_z}^2 + \gamma_e \Vec{B}_0 \cdot \Vec{\mathcal{S}},
\end{gather}
where $\mathcal{S}$ is the electron spin operator ($S = 1$), $D = 2870$~MHz is the zero-field splitting parameter at room temperature~\cite{cambria_physically_2023}.

\begin{figure*}[!ht]
	\centering \includegraphics{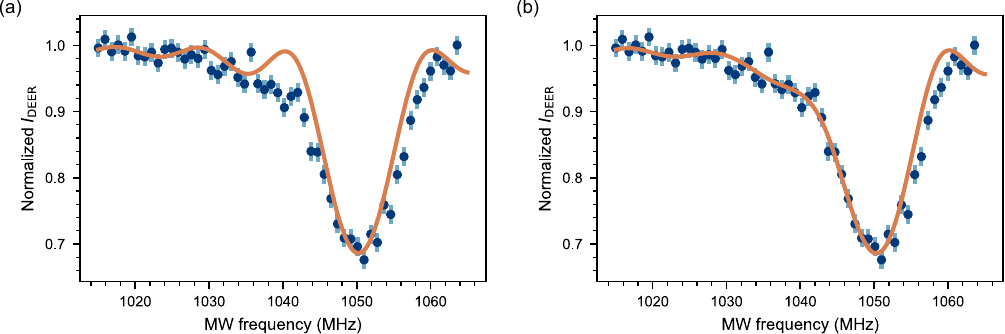}
	\caption{\label{fig:central_line} Experimentally measured (blue points) and numerically simulated using QuaCCAToo software~\cite{quaccatoo_paper, tsunaki_quantum_2024} DEER spectra (brown line) showing the central resonance line of P1 from Figure~\ref{fig:simulation}. (a) Simulations include only the P1 Hamiltonian (Equation~\ref{eq:p1_h}). (b) Simulations include both the P1 Hamiltonian and the defect X Hamiltonian (Equations~\ref{eq:p1_h} and~\ref{eq:x_h}) using $n^{}_\text{X} = 13$~ppb. Error bars represent the photon shot noise.}
\end{figure*}

The defect X (see Figure~\ref{fig:central_line}) was assumed to be a simple spin-$1/2$ system with the Hamiltonian
\begin{gather}
  \label{eq:x_h}
	\mathcal{H}_\text{X} = \gamma_e \Vec{B}_0 \cdot \Vec{\mathcal{S}},
\end{gather}
where $\mathcal{S}$ is the electron spin operator ($S = 1/2$).

The time-dependent Hamiltonian describing the interaction of the target spins with the DEER pulse was taken to be
\begin{gather}
  \label{eq:deer_pulse}
	\mathcal{H}_{1}(t) = \gamma_e \Vec{B}_1 \cdot \Vec{\mathcal{S}} \sin(2 \pi f_\text{B} t) = \Omega \: \Vec{e}_1 \cdot \Vec{\mathcal{S}} \sin(2 \pi f_\text{B} t).
\end{gather}
Here $\Vec{B}_1 = B_1 \Vec{e}_1$ is the MW magnetic field vector and $\Vec{e}_1$ is the unit vector showing its direction. In simulations we use $\Omega = \gamma_e B_1$, which is the experimentally determined DEER Rabi frequency of the detected spins.

In the simulations, one of the NV center and one of the P1 orientations (e.g. $[ 111 ]$) was chosen to be along the $z$-axis. The static magnetic field $\Vec{B}_0$ was placed in the $xy$-plane at an angle $\theta = 0.1$\textdegree{} to the $z$-axis to reproduce the NV DEER spectrum. The MW magnetic field $\Vec{B}_1$ was aligned with the $x$-axis. To simulate the NV and P1 centers of other crystallographic orientations ($[ \bar{1}11 ]$, $[ 1\bar{1}1 ]$, $[ 11\bar{1} ]$), a change of basis was applied: $\Vec{B'}_0 = R\Vec{B}_0$ and $\Vec{B'}_1 = R\Vec{B}_1$, where $R$ is the transition matrix describing the rotation of the $[ 111 ]$ vector to the new orientation (rotation around the y-axis by an angle $\theta_y = 109.5^{\circ}$ and consequent rotation around the z-axis by an angle $\theta_z = 0^{\circ},\ 120^{\circ},\ 240^{\circ}$).

DEER transitions were simulated as probabilities for changing the system initial state represented by a density matrix $\mathcal{\rho}_i = \ket{\psi_i}\bra{\psi_i}$ after application of the DEER pulse: $P_\text{B}^\text{sim} = 1 - \Tr(\rho_i \, \rho_f)$. Here, $\ket{\psi_i}$ are eigenstates of the Hamiltonian $\mathcal{H}_0 = \mathcal{H}_\text{P1} + \mathcal{H}_\text{NV} + \mathcal{H}_\text{X}$, representing the time-independent Hamiltonian of the simulated system (Equations~\ref{eq:p1_h},~\ref{eq:nv_h},~\ref{eq:x_h}). $\rho_f$ is the final density matrix of the system after DEER pulse. Finally, to compare simulations with the experimental data, we calculated $I^\text{sim}_\text{DEER}$ using Equation~\ref{eq:deer} with $P_\text{B}^\text{sim}$.

\begin{figure*}[!ht]
	\centering \includegraphics{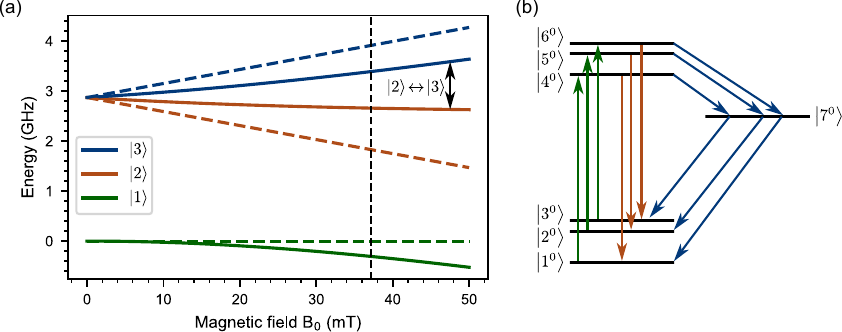}
	\caption{\label{fig:levels} (a) Ground state spin levels of the NV center aligned with $[ 111 ]$ crystallographic direction (dashed lines) and NV center of other orientations (off-axis NVs) as a function of the static magnetic field $B_0$ aligned at $\theta = 0.1$\textdegree{} to $[ 111 ]$. The magnetic field $B_0 = 37.2$~mT used in the simulations is shown by the black dashed line. (b) 7-state model of the NV center aligned with the magnetic field (zero perpendicular field). The arrows show the allowed transitions between states.}
\end{figure*}

For NV concentration measurements, we used the $ \ket{2} \leftrightarrow \ket{3} $ NV transition (see Figure~\ref{fig:levels}~(a)). To calculate $ |\sigma_\text{NV}| $ for this transition, we used eigenstates (in order of increasing energy) $\ket{1}$, $\ket{2}$, $\ket{3}$ of the NV Hamiltonian (Equation~\ref{eq:nv_h}) for one of the off-axis NV orientation (e.g. $[ \bar{1}11 ]$): $ |\sigma_\text{NV}| = \frac{1}{2} \left| \expval{\mathcal{S}_z}{2} - \expval{\mathcal{S}_z}{3} \right| = 0.87$.

To calculate the share of NV centers giving the observed signal in the DEER spectrum, we followed the method described in Ref.~\cite{li_determination_2021}. For that, the population of all states of the off-axis NVs was calculated using the 7 state model (see Figure~\ref{fig:levels}~(b)). Off-axis NV states $\ket{i}$ can be expressed as linear combinations of the states at zero perpendicular magnetic field $\ket{j^{0}}$: 
\begin{gather}
  \label{eq:states}
	\ket{i} = \sum_{j = 1}^{7} \alpha_{ij} \ket{j^{0}}.
\end{gather}
Coefficients $\alpha_{ij}$ were determined using the ground state Hamiltonian (Equation~\ref{eq:nv_h}) for $i, j = 1, 2, 3$ and the excited state Hamiltonian (Equation~\ref{eq:nv_h} with $D = 1420$~MHz) for $i, j = 4, 5, 6$. The singlet state $\ket{7} = \ket{7^{0}}$ was assumed to be not affected by the perpendicular magnetic field ($\alpha_{77} = 1$) and the rest of $\alpha_{ij}$ were taken to be zero.

The transition rates $k_{ij}$ between states $\ket{i}$ and $\ket{j}$ can then be expressed through the transition rates of the NV center at zero perpendicular magnetic field $k^{0}_{ij}$ shown in Figure~\ref{fig:levels}~(b):
\begin{gather}
  \label{eq:trrates}
	k_{ij} = \sum_{p = 1}^{7} \sum_{q = 1}^{7} |\alpha_{ip}|^2 |\alpha_{jq}|^2 k_{pq}^{0},
\end{gather}
where $k_{41}^{0} = k_{52}^{0} = k_{63}^{0} =  65.9$~\textmu{s}$^{-1}$, 
$k_{14}^{0} = k_{25}^{0} = k_{36}^{0} = \beta k_{41}^{0}$, 
$k_{47}^{0} = 7.9$~\textmu{s}$^{-1}$,
$k_{57}^{0} = k_{67}^{0} =  53.3$~\textmu{s}$^{-1}$,
$k_{71}^{0} = 1.0$~\textmu{s}$^{-1}$,
$k_{72}^{0} = k_{73}^{0} =  0.7$~\textmu{s}$^{-1}$ taken as average values for the rates derived in Reference~\cite{tetienne_magnetic-field-dependent_2012} for 4 NV centers with different orientations. The coefficient $\beta$ related to the laser pumping power was taken to be $0.03$~\cite{li_determination_2021}.

The populations of the NV states $n_{i}$ can then be obtained solving a system of the rate equations:
\begin{gather}
	\frac{\dd n_i}{\dd t} = \sum_{j = 1}^{7} (k_{ji} n_j - k_{ij} n_i), \notag \\
  \sum_{i = 1}^{7} n_i = 1.
  \label{eq:rateeqs}
\end{gather}

Note, that since the coefficient $\beta$ does not affect the rates of spin-mixing transitions, the maximum achievable spin polarization does not depend on it. In the experiments, the time between two consecutive laser pulses was $\approx 160$~\textmu{s}, which is an order of magnitude less than the typical NV spin relaxation time in IIa crystals even at misaligned magnetic fields~\cite{mrozek_longitudinal_2015}. Therefore, we can neglect the spin relaxation during the measurement. Moreover, since we use the $ \ket{2} \leftrightarrow \ket{3} $ NV transition, which have the same population due to the fact that their transition rates are the same, the population of the spin states is not changed during the MW $\pi$-pulse. Thus, subsequent laser pulses will eventually bring the system to a steady state with a maximum spin polarization.

Taking the transition rates mentioned above, the laser length pulse of 5~\textmu{s}, the time between the laser pulses 160~\textmu{s}, the waiting time after the last laser pulse of 1.5~\textmu{s}, and considering equal population of the ground state spin levels as initial conditions ($n_i(t = 0) = 1/3$ for $i = 1, 2, 3$ before the first laser pulse) we calculate the population of spin levels before MW application to be $n_1 = 40$\% and $n_2 = n_3 = 30$\%, which is reached already after 1--15 laser pulses for $\beta$ in range $[0.001, 0.1]$. Therefore, the share of NV centers undergoing the $\ket{2} \leftrightarrow \ket{3}$ transition is $n_2 + n_3 = 60$\% of the off-axis NV centers. As these account for 75\% of the total amount of NV centers, in total 45\% of all NVs give the observed signal in DEER measurements.


\clearpage
\widetext

\setcounter{equation}{0}
\setcounter{figure}{0}
\setcounter{table}{0}
\setcounter{page}{1}
\makeatletter
\renewcommand{\theequation}{S\arabic{equation}}
\renewcommand{\thefigure}{S\arabic{figure}}
\renewcommand{\thesubsection}{S\arabic{subsection}}
\renewcommand{\thepage}{S\arabic{page}}

\section*{\Large Supplementary Material}

\subsection{Electron paramagnetic resonance measurements}
\begin{figure}[!ht]
	\centering \includegraphics[width=\columnwidth]{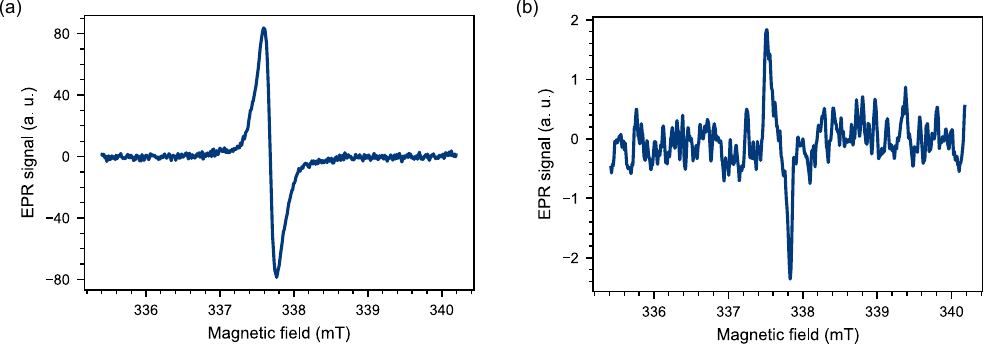}
	\caption{\label{fig:epr} EPR spectra of (a) reference sample and (b) studied sample used for estimation of the average P1 concentration.} 
\end{figure}

The P1 concentration in the studied sample (mass $m = 11$~mg) was estimated using quantitative electron paramagnetic resonance (EPR)~\cite{eaton_quantitative_2010}. This method is based on the analytical expression of the double integral $DI$ of the EPR spectrum, which is proportional to the number of spins. Using a reference sample with a known number of spins, it is possible to determine the spin concentration in the sample of interest. In this work we used a diamond sample (mass $m^\text{ref} = 50.7$~mg) with known amount of P1 centers ($n^\text{ref}_\text{P1} = 68$~ppm) as a reference. The analysis revealed the average nitrogen concentration $n^{}_\text{P1} = 22$~ppb in the studied sample. The EPR spectra used to determine the P1 center concentration are shown in Figure~\ref{fig:epr}.

\clearpage

\subsection{SRIM simulations}

\begin{figure}[!ht]
	\centering \includegraphics{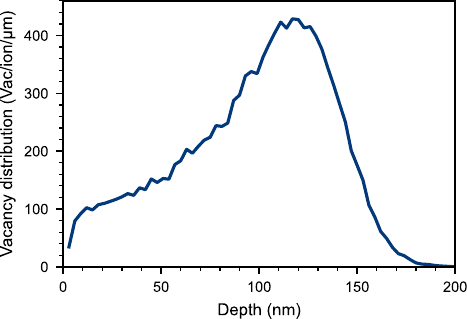}
	\caption{\label{fig:srim} Vacancy depth distribution calculated using SRIM.} 
\end{figure}

The ion implantation was simulated using the Stopping and Range of Ions in Matter (SRIM) software~\cite{ziegler_srim_2010}. The parameters used in the simulations are: $^4$He$^+$ ion energy --- 30~keV, angle of incidence --- $0^{\circ}$, diamond density --- 3.53~g/cm$^3$, carbon atom displacement energy --- 40~eV, lattice binding energy --- 3~eV, surface binding energy --- 7.41~eV. The resulting depth distribution of vacancies is shown in Figure~\ref{fig:srim}. The vacancy distribution radius used in the main text is extracted from the full width at half maximum (FWHM): $r^{}_\text{vac} = \text{FWHM}/2=37.5$~nm.

\clearpage


\subsection{Hahn echo decay measurements}

\begin{figure}[!ht]
	\centering \includegraphics[width=\columnwidth]{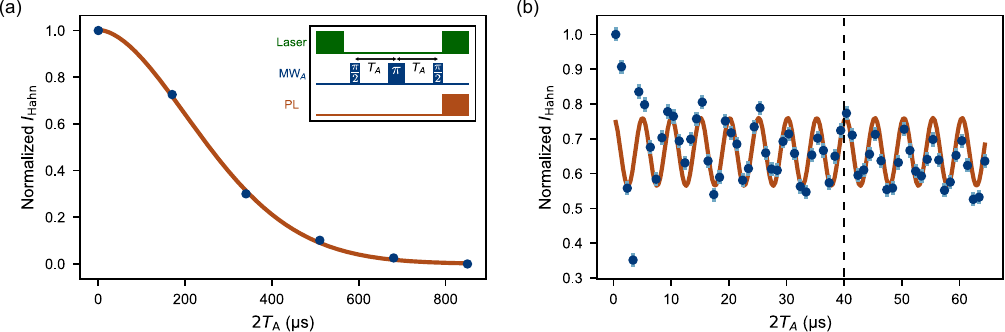}
	\caption{\label{fig:hahn} Hahn echo decay measurements conducted at a 5000 ions/spot implantation site in $\{ 111 \}$ GS. (a) Hahn echo decay in a large range of $T_\text{A}$. The data (blue points) is fitted with Equation~\ref{eq:hahn1} (brown line). Parameters obtained from the fit: $T_2 = 313 \pm 5$~\textmu{s}, $n = 1.80 \pm 0.07$. Inset shows the pulse sequence used in the experiment. (b) ESEEM. The data (blue points) is fitted with Equation~\ref{eq:hahn2} (brown line). Parameters obtained from the fit: $f = 198.5 \pm 0.5$ kHz. The dashed line indicates the value of $T_\text{A} = 20$~\textmu{s}, used in DEER measurements. Error bars represent the photon shot noise.} 
\end{figure}

Hahn echo decay was measured to determine the coherence time $T_2$ of the NV ensembles and to set the $T_\text{A}$ value for the DEER pulse sequence. The $T_2$ values were obtained by fitting the data $I_\text{Hahn}$ with a stretched exponential function (see Figure~\ref{fig:hahn}~(a)):
\begin{gather}
	\label{eq:hahn1}
	I^{}_\text{Hahn}(2T_\text{A}) = e^{-\left(\frac{2T_\text{A}}{T_2}\right)^n}
\end{gather}

On small scales of $T_\text{A}$, the measurements reveal a sinusoidal modulation of $I_\text{Hahn}$ (electron spin echo envelope modulation, ESEEM) shown in Figure~\ref{fig:hahn}~(b), induced by coupling to the $^{13}$C nuclear spin bath~\cite{van_oort_optically_1990}. To obtain the modulation frequency, the data can be fitted with

\begin{gather}
	\label{eq:hahn2}
	I^{}_\text{Hahn}(2T_\text{A}) = \cos{(2 \pi f \, 2 T_\text{A})} + C
\end{gather}

The frequency $f$ obtained from the fit can be used to calculate the gyromagnetic ratio $\gamma_n$ of the detected nuclear spins according to the formula $\gamma_n = 2f/B_0$, where $B_0$ is the static magnetic field. For the data shown in Figure~\ref{fig:hahn}~(b), $\gamma_n = 10.68 \pm 0.03$~MHz/T, which agrees well with the expected gyromagnetic ratio for $^{13}$C  $\gamma_{n}(^{13}\text{C}) = 10.71$~MHz/T.

\clearpage


\subsection{Photoluminescence saturation measurements}\label{sec:satcurve}

To determine the number of NV centers created in the implantation sites, PL saturation was measured. During the experiments, the PL of the implanted spots, of the single NVs and of the background was measured as a function of the laser power (measured before the objective) at zero magnetic field. To account for the fact, that at high photon count rates the avalanche photodiode (APD) is working in a nonlinear regime, the PL intensity measured on NV ensembles was multiplied by a correction factor specified in the APD manual.

\begin{figure}[!ht]
	\centering \includegraphics[width=\columnwidth]{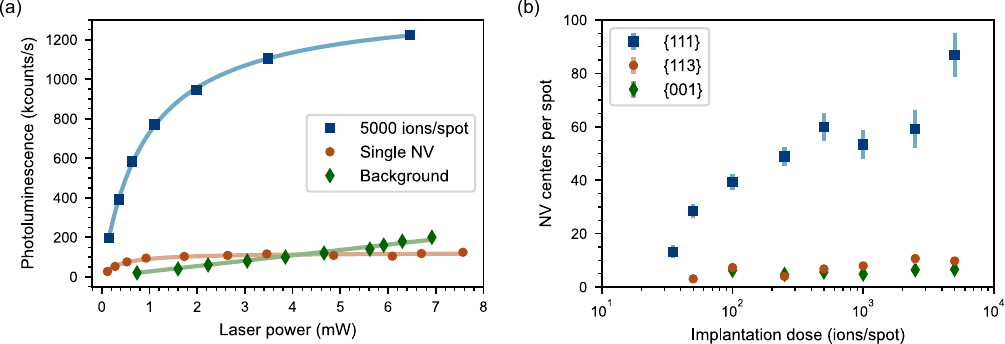}
	\caption{\label{fig:sat} (a) Saturation curves for the implanted spot with a dose of 5000 ions/spot (blue squares), for the single NV (brown circles) and for the background (green diamonds) in $\{ 113 \}$ GS. Solid lines of lighter colors represent the fitting curves. (b) The number of NV centers as a function of the implantation dose for $\{ 111 \}$ GS (blue squares), for $\{ 113 \}$ GS (brown circles), and for $\{ 001 \}$ GS (green diamonds). Error bars represent the standard errors obtained from fitting.} 
\end{figure}

As explained in the main text, by fitting the resulting saturation curves for implanted sites and for single NVs, one can estimate the number of NV centers in spots with NV ensembles. Examples of saturation curves for $\{ 113 \}$ GS and for 5000 ions/spot implantation dose are shown in Figure~\ref{fig:sat}~(a). The dependencies of the amount of NV centers on the implantation dose for all three GSs are presented in Figure~\ref{fig:sat}~(b).

\clearpage


\subsection{Photoluminescence maps}

\begin{figure}[!ht]
	\centering \includegraphics[width=\columnwidth]{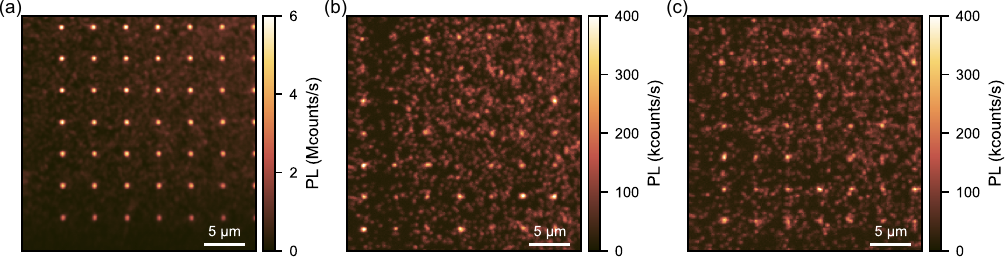}
	\caption{\label{fig:pl_maps} PL maps obtained in different GSs of the sample showing spots with implantation doses increasing in the vertical direction from top to bottom: 500, 1000, 1000, 2500, 2500, 5000, 5000 ions/spot. In the horizontal direction the doses are not changing. (a) PL map of the implanted region in $\{ 111 \}$ GS. The bottom of the map is darker due to the shadow of the copper wire used for application of MWs. (b) PL map of the implanted region in $\{ 113 \}$ GS. (c) PL map of the implanted region in $\{ 001 \}$ GS.} 
\end{figure}

After the ion implantation and annealing, the NV centers appeared not only in the implantation spots, but also between them. Figure~\ref{fig:pl_maps} shows photoluminescence (PL) of the implanted regions from all three growth sectors (GS), where these additional NVs (some even on a single defect level) can be resolved. These single NV centers were used for saturation curve measurements (see Section~\ref{sec:satcurve}). The exact origin of these single NV centers is still not clear.

\end{document}